\documentclass[preprint]{aastex}
\usepackage[]{natbib}
\begin{document}

\title{Probing Shock Properties with Non-thermal X-ray Filaments in Cas A}

\author{Miguel Araya\altaffilmark{\dag,1}, David Lomiashvili\altaffilmark{\dag,\S,2}, Chulhoon Chang\altaffilmark{\dag,3}, Maxim Lyutikov\altaffilmark{\dag,4} and Wei Cui\altaffilmark{\dag,5}
\\
\scriptsize{$^\dag$ Department of Physics, Purdue University, West Lafayette, USA}
\\
\scriptsize{$^\S$ E. Kharadze Georgian National Astrophysical Observatory, Ilia Chavchavadze State University, Tbilisi, Georgia} }
\altaffiltext{1}{marayaar@purdue.edu}
\altaffiltext{2}{dlomiash@purdue.edu}
\altaffiltext{3}{Now at the Pennsylvania State University, PA, USA; chchang@astro.psu.edu}
\altaffiltext{4}{lyutikov@purdue.edu}
\altaffiltext{5}{cui@purdue.edu}

\begin{abstract}
Thin non-thermal X-ray filaments are often seen in young supernova remnants. We used data from the 1 Ms \emph{Chandra} observation of Cassiopeia A to study spectral properties of some of the filaments in this remnant. For all the cases that we examined, the X-ray spectrum across the filaments hardens, at about 10\% level, going outward, while observed filament widths depend only weakly on the photon energy. Using a model that includes radiative cooling, advection and diffusion of accelerated particles behind the shock, we estimated the magnetic field, turbulence level, and shock obliquity.
\end{abstract}


\section{Introduction}
Young supernova remnants (SNRs) have long been thought to be the main source of galactic cosmic rays \citep{shk53}. Evidence for the existence of high-energy electrons in SNRs first came with the detection of non-thermal emission in the radio and later in X-rays \citep[e.g.,][]{koy95,bam00,sla01}. With the use of the \emph{Chandra X-Ray Observatory}, detailed images of SNRs have revealed very thin structures \citep{lon03} near the forward shock. The spectral and spatial properties of such structures, often referred to as filaments, are consistent with synchrotron emission from highly relativistic electrons. High-energy protons and nuclei are also believed to be produced within SNRs, although no direct evidence has been conclusively found. However, a recent analysis of the \emph{Fermi}-LAT spectrum of the SNR W51C suggests that the main component of emission in the GeV band from this object is produced through interactions of high-energy hadrons \citep{fer09}.

In SNRs, charged particles may gain energy by repeatedly crossing the shock \citep{bel78II,bla78,dru83}. The process is thought to be facilitated by scattering off magnetic turbulences downstream and magnetic irregularities upstream \citep{bla87}. Such diffusive shock acceleration (DSA) naturally leads to a power-law distribution of particles, which is seen for cosmic rays. However, little is known about the precise nature of the magnetic turbulences. The scattering of particles by circularly polarized MHD waves with a frequency equal to the gyrofrequency of the particles results in diffusion \citep{bla87}. Quantifying the magnetic diffusion, therefore, offers a way to probe the turbulent wave spectrum in SNRs.

Many issues regarding DSA are still unresolved \citep[for a summary, see][and references therein]{rey08}. For instance, some of the observed electron spectral distributions have indices that are either too high or too low to be accounted for by the mechanism of DSA alone and may require that other effects be considered, such as nonlinear processes (for example, the deceleration of the incoming fluid with respect to the shock \citep{eic79} or the excitation of MHD waves by the accelerated particles themselves \citep{bel78I}), as well as the effects of second-order Fermi acceleration \citep{ost93}.

The injection problem is another unresolved issue. One of the requirements of DSA is that particles have enough initial energy to be able to pass through the shock without being significantly deflected and, therefore, that their gyroradii be greater than the shock thickness. Since the electron gyroradii in the thermal plasma of SNRs are typically smaller than the shock thickness, this poses a problem for explaining the initial acceleration mechanism for these particles \citep[for a discussion on this problem, see][]{mal01}.

The understanding of DSA requires knowledge of magnetic field. There is evidence for magnetic field amplification in SNRs \citep{ber02,vink03,ber04} with respect to pre-existing ambient field. A mechanism for explaining such amplification has been treated for the case where the magnetic field is parallel to the shock normal \citep[i.e., with an obliquity angle of zero, see][]{bel01}, although much is still unknown about the possible dependence of this amplification with obliquity angle, shock speed, or composition of the upstream medium.

Another feature of shock acceleration that has been debated is the so-called shock precursor, a region supposedly formed by scattering of high-energy particles upstream of the shock. Since DSA requires that particles cross the shock back and forth it is possible that particles with very high energy scatter in a region ahead of the shock. It is expected that the scale of the precursor will depend on properties such as the particle density and the diffusion coefficient. The existence of such a precursor is still debated and its properties are largely unknown \citep{ell94}.

To gain insights into some of these unresolved issues we studied X-ray synchrotron emission from Cassiopeia A (Cas A), a young SNR with an age of approximately 300 yr \citep{hug80}. Much has already been learned about this remnant, from the identification of the forward and reverse shocks \citep{got01} to the measurements of the magnetic field \citep{vink03} as well as the observed proper motion of the forward shock \citep{del03,del04,pat09a}. The expansion rate of the SNR seems to be lower than expected, due probably to more efficient particle acceleration \citep{pat09a} or to a more complicated density profile of the ambient medium  \citep{hwa09}. As will be seen, our analysis does not allow us to draw conclusions regarding these possibilities, partly because it only deals with the leptonic component of cosmic rays in Cas A, and partly because we do not consider the dynamics of the expansion.

In this work, we used data from the 1 Ms exposure on Cas A with \emph{Chandra} \citep{hwa04} to carry out a detailed spectral analysis of non-thermal filaments in the outermost region of the remnant. Specifically, we are interested in the energy dependence of the width of the filaments and spectral variation across them. In the context of advection and synchrotron radiative cooling, the widths are expected to decrease with increasing energy and the spectrum softens going downstream. However, diffusion may significantly modify the behaviors.

\section{The Observation}
We extracted spectra from nine non-thermal filaments of Cas A from the archival \emph{Chandra} 1 Ms observation, taken with the backside-illuminated S3 CCD chip of the Advanced CCD Imaging Spectrometer (ACIS). The observation consists of nine segments, the first was taken in 2004 February and the rest between 2004 April and May \citep[see][]{hwa04}. For our analysis we used level 2 archival data products and reduced the data with the standard \emph{Chandra} software package CIAO v3.4 and \emph{Chandra} calibration database (CALDB) version 3.5.2. Data were accumulated in GRADED mode to avoid telemetry loss, therefore the effects of charge transfer inefficiency in the spectra cannot be corrected. This loss of charge affects the measured pulse-height distribution and the energy resolution, although the effect is small for the backside-illuminated CCDs \citep{tow00}.

\begin{figure}[htp]
\centering
\includegraphics[width=14cm,height=14cm]{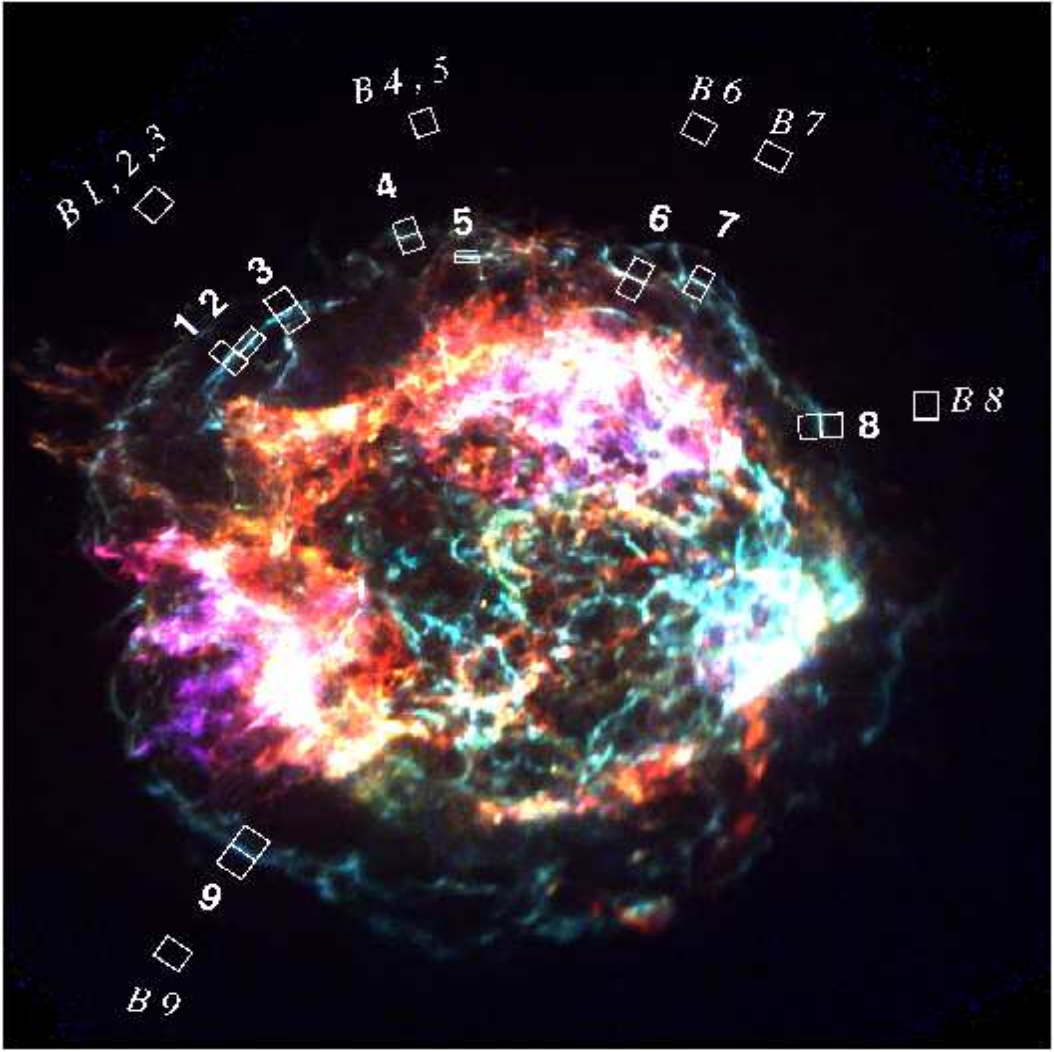}
\caption{Exposure-corrected image of Cas A in the range from 0.3 keV to 10 keV. The selected non-thermal filaments are indicated, along with source and background regions for spectral extraction. \label{fig-1}}
\end{figure}

The image of Cas A, shown in Figure \ref{fig-1}, was obtained by combining events in the energy range from $0.3 - 10$ keV and then correcting the count map by effective exposures. Since the effective area is energy dependent, weighted exposure maps were calculated at different energies and combined. We focused on regions that had previously been identified as being non-thermal \citep{sta06} and been thought to be associated with the forward shock \citep{got01}. However, due to low statistics (even with a 1 Ms exposure), faint filaments mainly located in the western and eastern sides of the remnant were not included in the analysis. The image also shows the nine filaments chosen for this work along with off-source regions for background estimates.

\subsection{Dividing the filaments}
Each of the non-thermal filaments was divided into an `inner' and an `outer' region, with the `inner' region being closer to the interior of the remnant. The division between the inner and outer regions for each filament was set at the peak of its linear intensity profile. This division made it possible for us to quantify the difference in the spectral properties of the radiation emitted by the electrons at different locations. One may naively attribute the difference to the fact that electrons in the inner region have had more time to evolve after interacting with the forward shock than the electrons in the outer region. However, as we will show, there seems to be a fair amount of mixing, implied by the inferred diffusion coefficients.

\begin{figure}[htp]
\centering
\includegraphics[width=10cm]{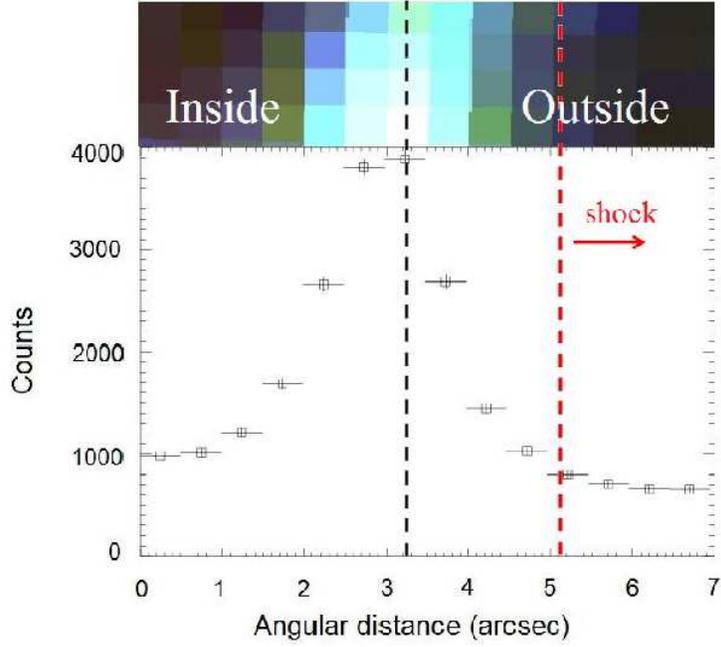}
\caption{Linear intensity profile of Filament 5 in the energy range 0.3 keV-10 keV. The top panel shows a \emph{Chandra} image of the filament with the inner and outer regions labeled and the position of the shock indicated. \label{fig-2}}
\end{figure}

As an example, Figure \ref{fig-2} shows the linear profile of Filament 5 (for the $0.3 - 10$ keV band). This profile was obtained from a $1^{\prime\prime}.5 \times 6^{\prime\prime}.9$ region running perpendicular across the filament with a bin size of $0^{\prime\prime}.5$. The top panel shows the division between the inner and outer emitting zones.

\subsection{Energy dependence of filament widths}
\begin{figure}[htp]
\centering
\includegraphics[width=10cm]{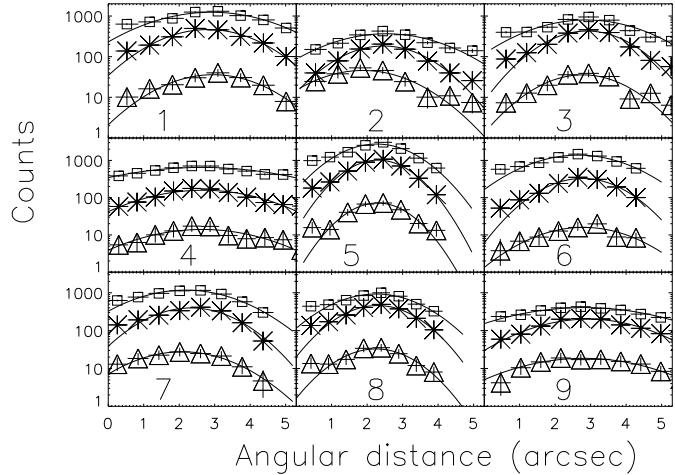}
\caption{Linear profiles of the filaments near the peak, for three energy bands (from top to bottom): $0.3 - 2$ keV, $3 - 6$ keV, and $6 - 10$ keV. The solid lines show the best-fit Gaussian functions. \label{fig-3}}
\end{figure}

The widths of each filament were estimated in three energy bands: $0.3 - 2$ keV, $3 - 6$ keV and $6 - 10$ keV. In order to quantify the width of a linear profile, we fitted the profile around the peak with a Gaussian function, as shown in Figure \ref{fig-3}. The results of the fits are summarized in Table \ref{tbl-1}. No strong dependence of the widths on energy is apparent.

We should note that the overall linear profiles of the filaments are highly non-Gaussian (see, e.g., Figure \ref{fig-2}). Nevertheless, we think that the derived Gaussian widths reflect fairly accurately the widths of the filaments.

\subsection{Spectral evolution of filaments}
To carry out the spectral analysis, we reprojected the event 2 files to a common tangent point and used the CIAO tool \emph{acisspec} to extract events between 0.3 keV and 10 keV from each region shown in Figure \ref{fig-1}, calculate weighted Auxiliary Response Files, and combine the spectra from the individual segments of the observation. Consistent calibration was used separately to produce the exposure-weighted responses by applying time-dependent gain corrections appropriate for $-120^{\circ}{\rm C}$ GRADED mode data on the back-illuminated S3 chip.
After the individual files were combined, we binned each spectrum such that each bin contained at least 100 counts and proceeded to individually model them with XSPEC version 11.3.2 \citep{arn96}.
\par
\begin{figure}[htp]
\centering
\includegraphics[width=5.4cm,height=4.6cm]{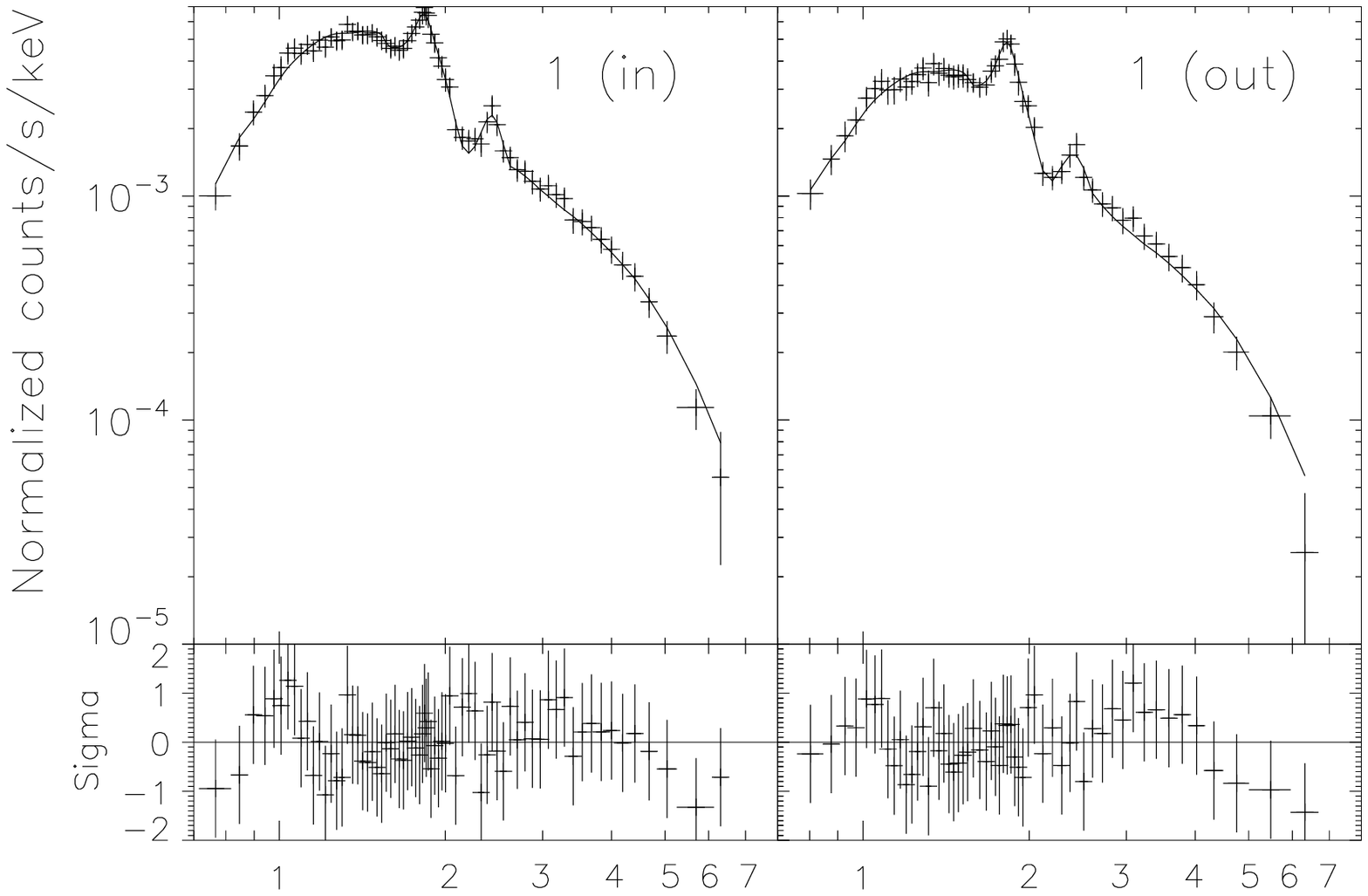}
\includegraphics[width=5.4cm,height=4.6cm]{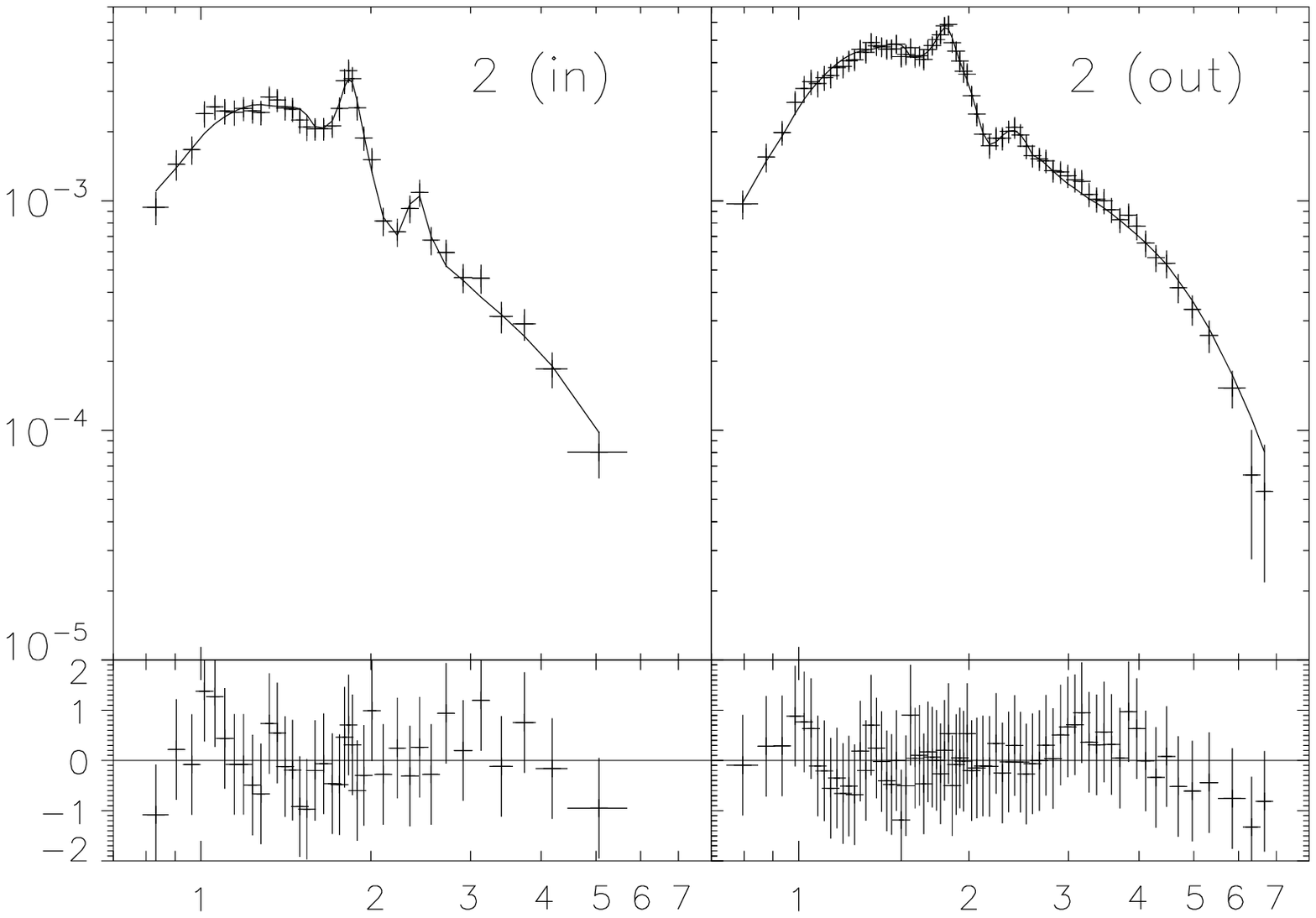}
\includegraphics[width=5.4cm,height=4.6cm]{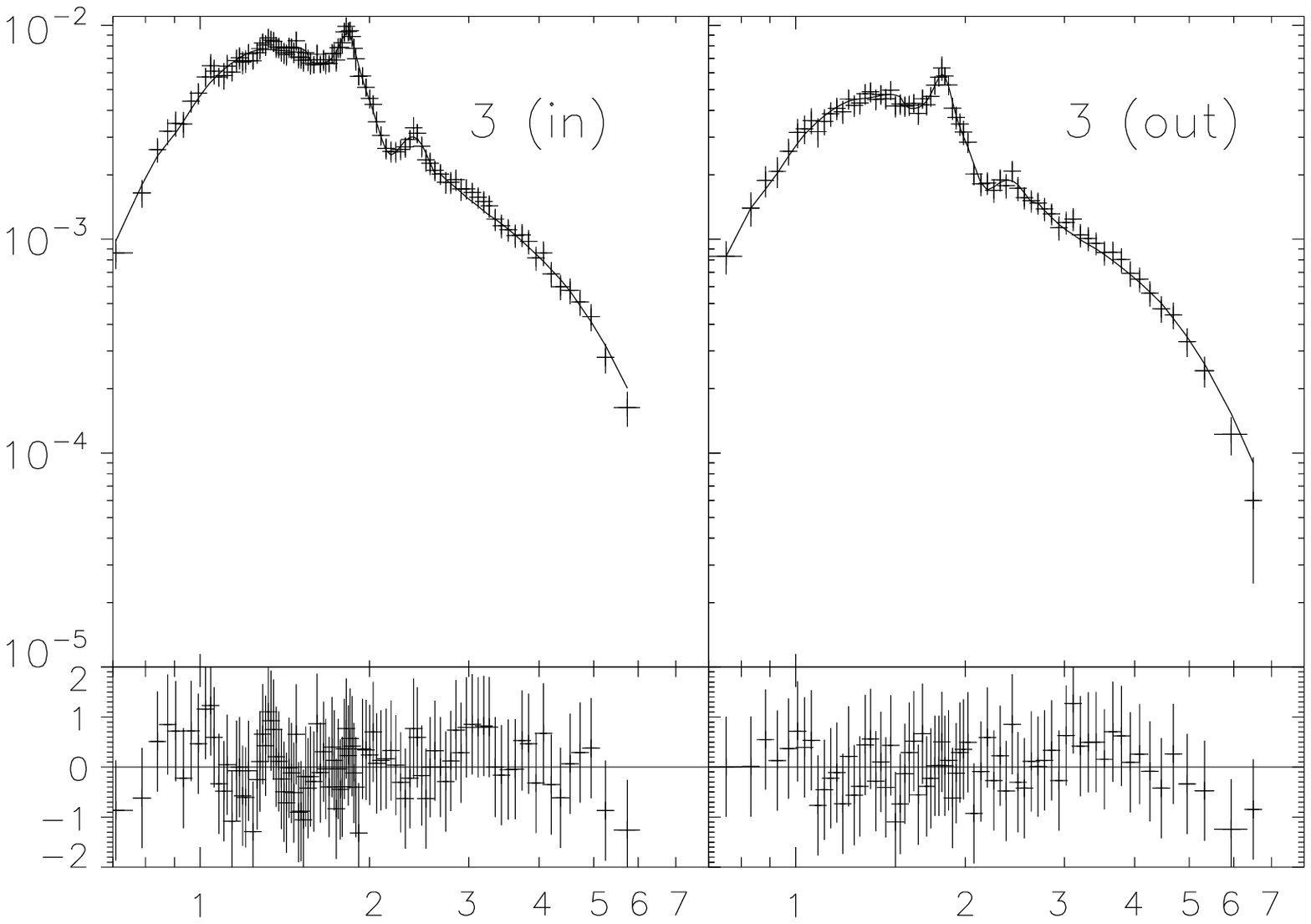}
\includegraphics[width=5.4cm,height=4.6cm]{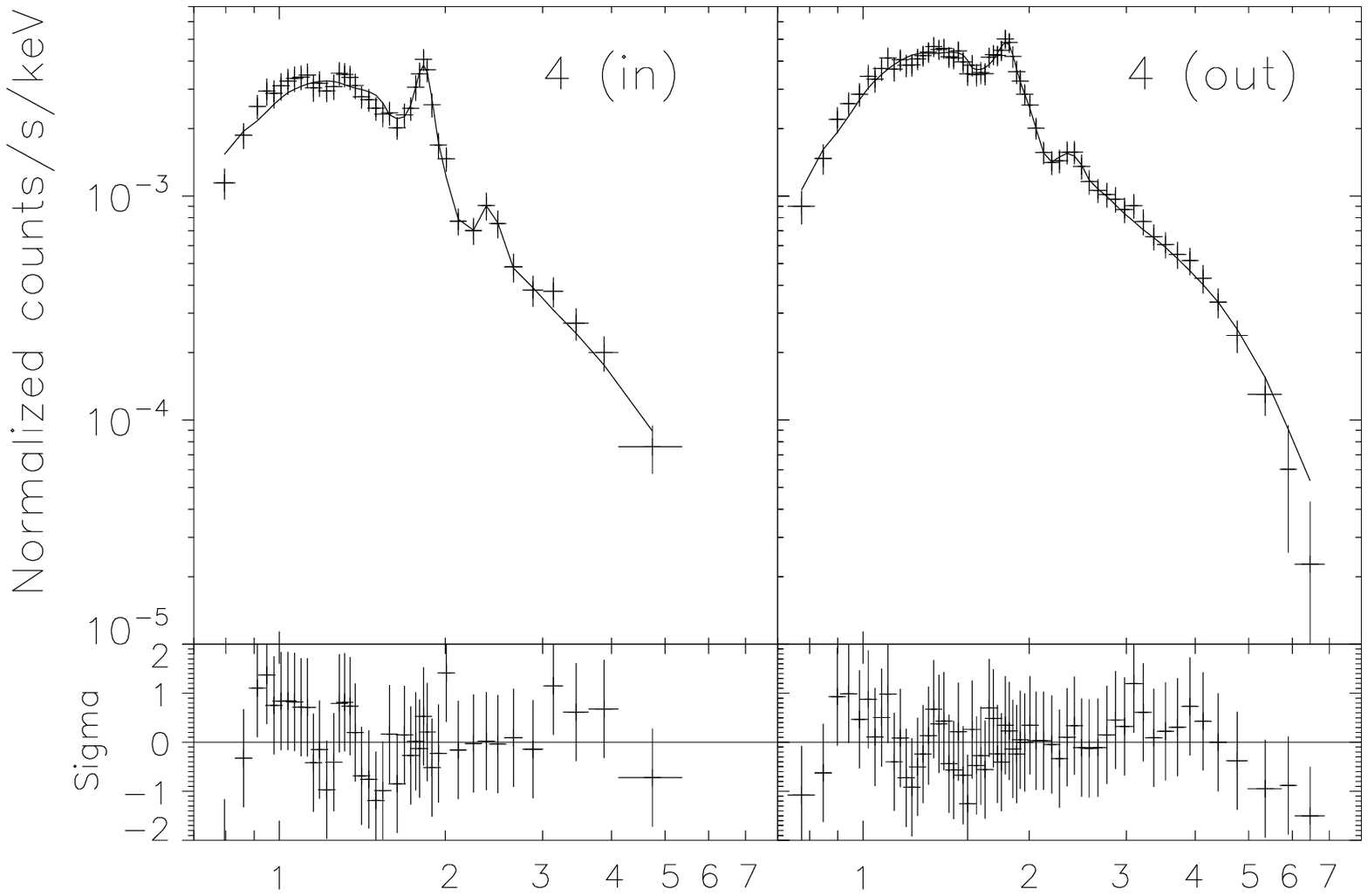}
\includegraphics[width=5.4cm,height=4.6cm]{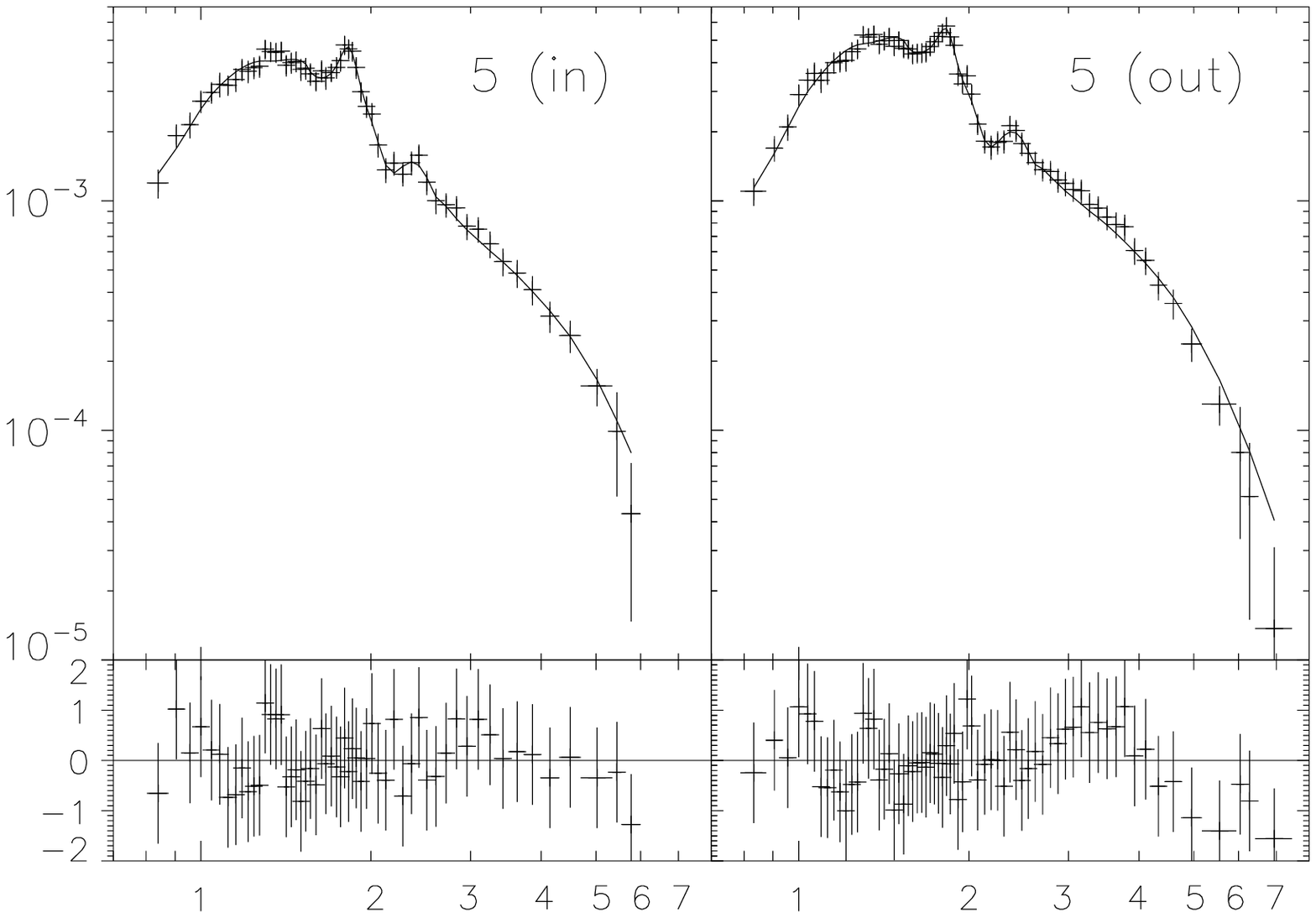}
\includegraphics[width=5.4cm,height=4.6cm]{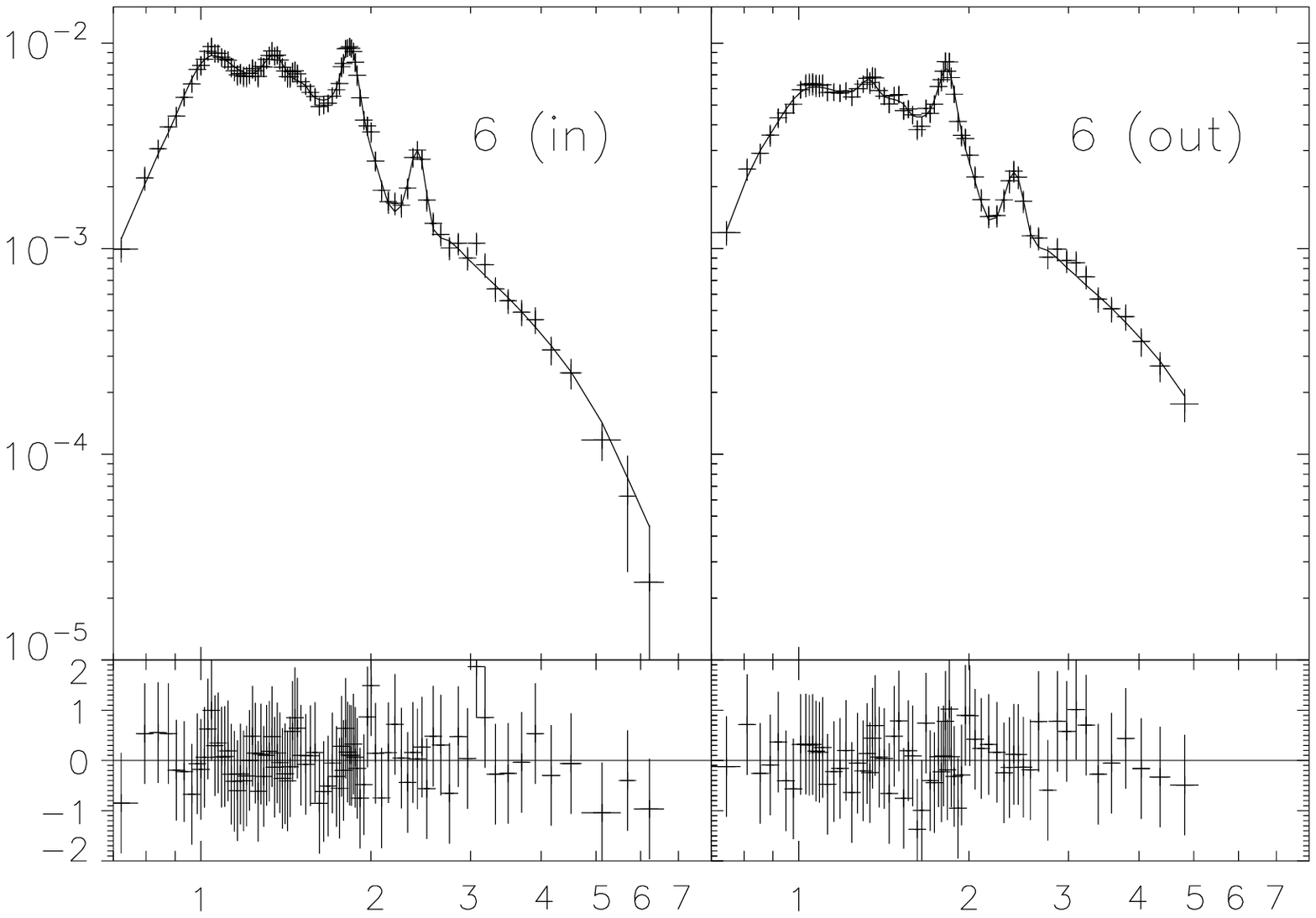}
\includegraphics[width=5.4cm,height=4.6cm]{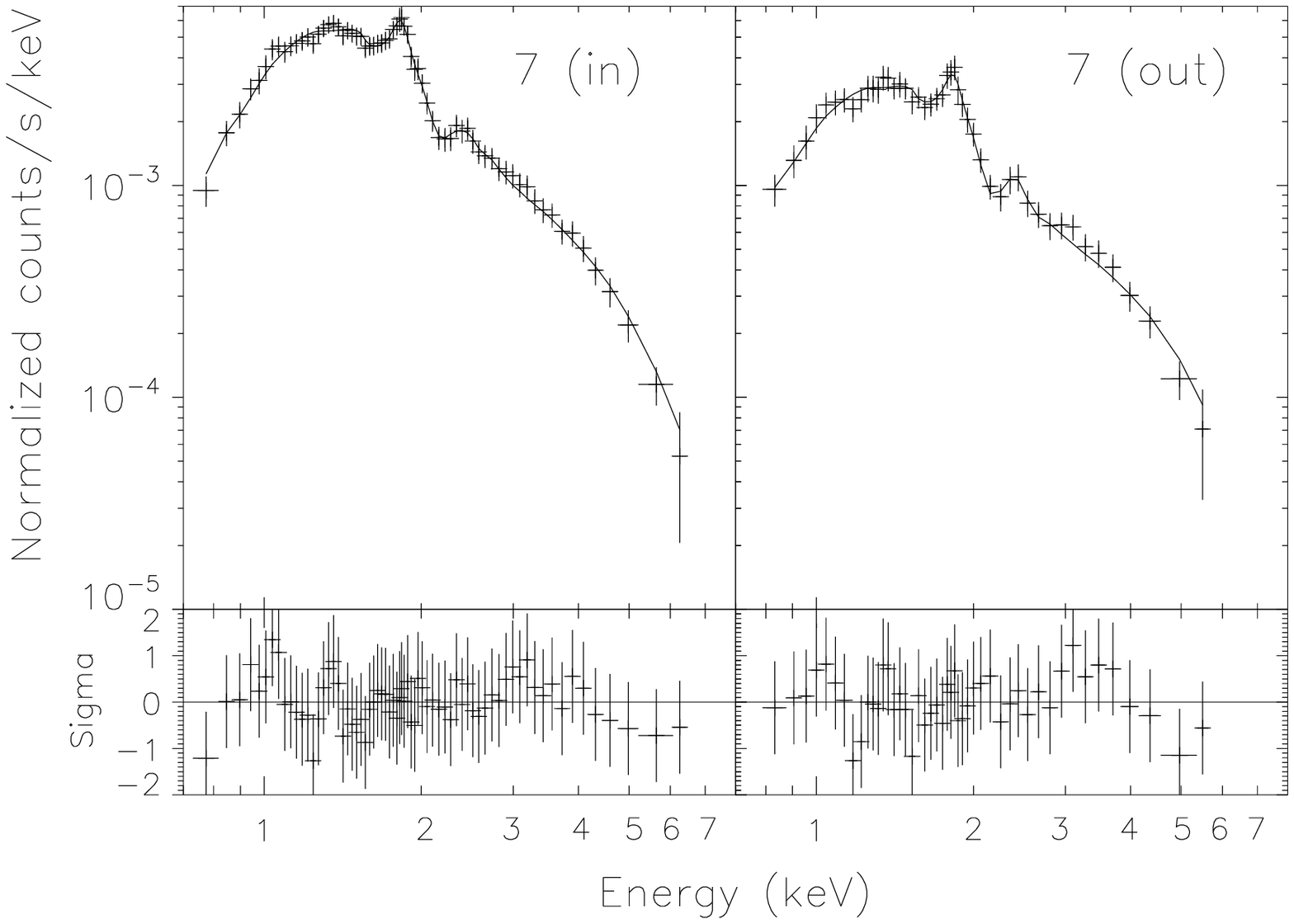}
\includegraphics[width=5.4cm,height=4.6cm]{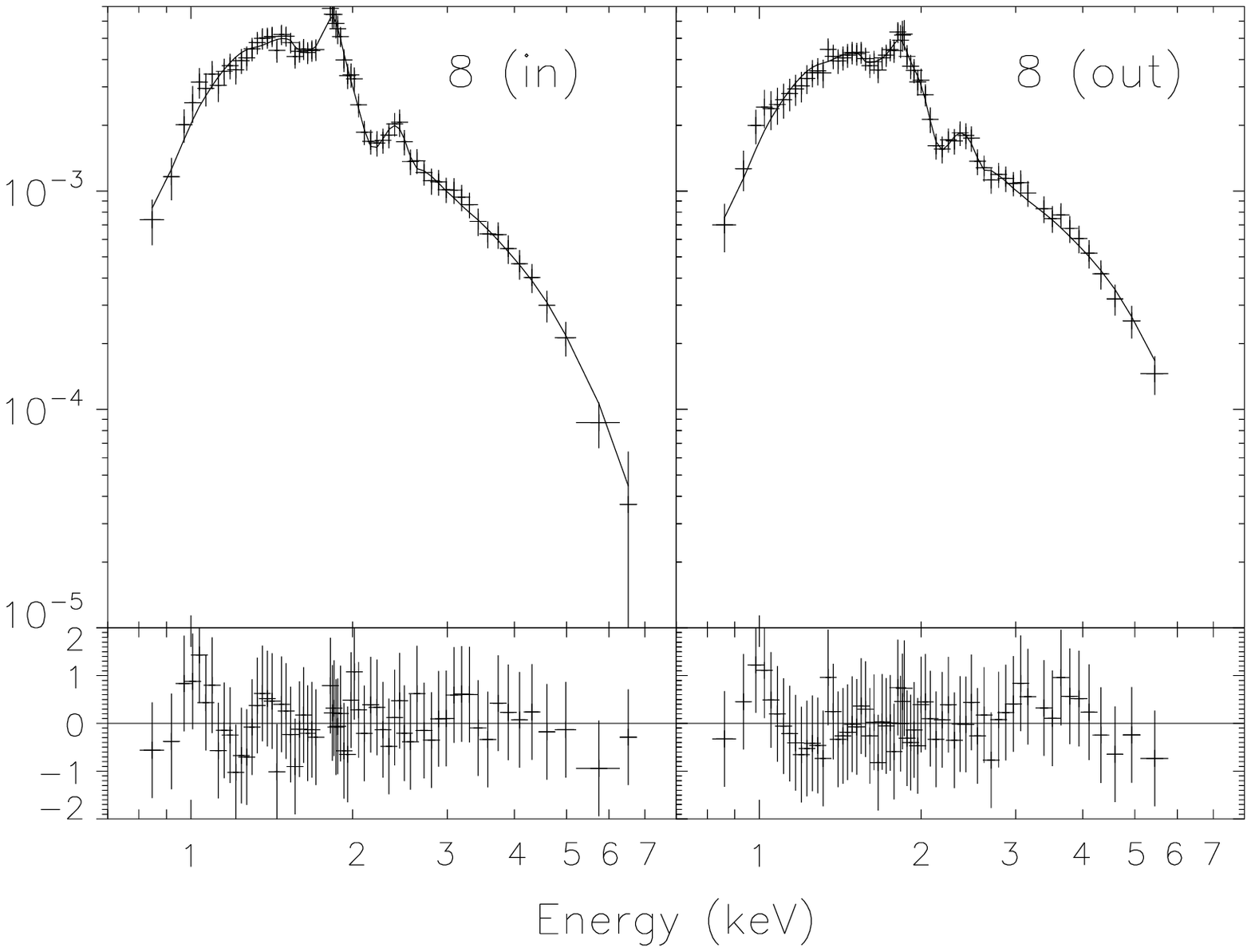}
\includegraphics[width=5.4cm,height=4.6cm]{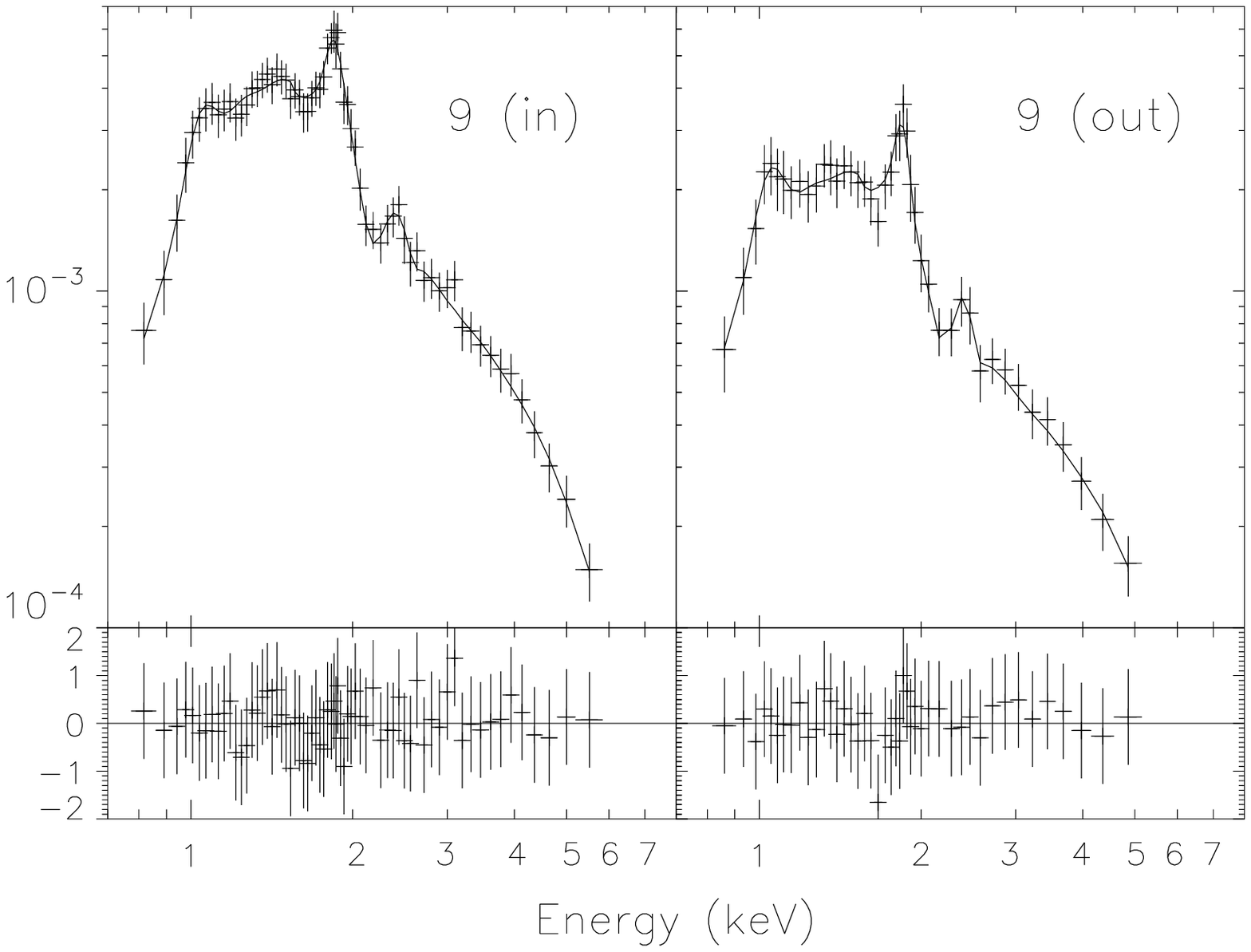}
\caption{Spectra of the inner and outer regions of non-thermal filaments. The solid lines show the best power-law fits to the spectra. The residuals of the fits are also shown. \label{fig-4}}
\end{figure}

All spectra show, with varying degrees of prominence, the presence of emission lines (see Figure \ref{fig-4}), indicating the existence of thermal photons in the regions. For most cases, we added two Gaussian components to model the lines at around 1.85 and 2.38 keV, which we attribute to Si K XIII and S K XV, respectively. Other weaker lines also appear to be present in some filaments. Filaments 5 and 6 show two additional lines at 1.3 keV and 1.0 keV, most likely associated with Mg XI and Fe XXI, respectively. Filament 9 also shows the line at 1.0 keV.

The inner and outer extraction regions have a typical extension of about $7^{\prime\prime}$ each, which corresponds to a physical size of roughly 0.1 pc (assuming a distance of 3.4 kpc to the remnant; \citep[see][]{ree95}). The regions for Filaments 2 and 5 are smaller ($4^{\prime\prime}$ and $3^{\prime\prime}$, respectively) since there seems to be a considerable amount of thermal emission in these areas. To assess possible ``contamination" from thermal emission, we also experimented with thinner extraction regions for each filament as well as on-source background regions. In the first case, we failed to remove the lines seen, while in the second one it becomes difficult to determine the appropriate locations of background regions necessary to avoid subtraction of non-thermal photons. The resulting lack of statistics after the subtraction generally does not allow to carry out a satisfactory analysis of the non-thermal X-rays. It is possible that the thermal and non-thermal emissions are cospatial, but we do not rule out that the detection of thermal photons might be due (at least partly) to scattered X-rays.

The photon spectra of all filaments were satisfactorily fitted with an absorbed power-law, with indices ranging from 2.2 to 3. Figure \ref{fig-4} shows the spectral fits, as well as residuals, for both the inner and outer regions of each filament. The results are summarized in Table \ref{tbl-2}.

Although in most cases the error intervals of the photon indices for the inner and outer regions overlap, we note that the spectrum of the inner region is in general softer than that of the outer one. The difference in photon indices between the inner and outer regions is on the order of 10\%.

The hydrogen absorption column values obtained from the fits are typically $0.7 - 0.9 \, \times{10}^{22}\,$cm$^{-2}$ in all regions except for a larger value of $1.3\times10^{22}\,$cm$^{-2}$ found for Filament 8, at the western edge of the remnant, where it is believed that it is interacting with a molecular cloud \citep{keo96}.

\section{Theoretical modeling}
We developed a model to explain the observational results. The model takes into account synchrotron radiative losses and diffusion of particles in the forward shock region. We assumed that the injected particles follow a power-law spectral distribution with index $\Gamma$ and the particle spectrum subsequently evolves.

We approximated Cas A as a sphere with radius $R=10^{19}\,$cm \citep{ree95}. The non-thermal emission is assumed to come from a thin shell near the edge of the sphere and integrated along the line of sight. The evolution of the non-thermal electron distribution is given by the diffusion-loss equation. We used the solution derived by Syrovatskii (1959), but also included the advection process. For Cas A, the advection speed of the plasma downstream of the shock is $V_{\mbox{adv}}=1300 \,$ km s$^-1$, equal to the shock speed $V_{\mbox{sh}}$ divided by a shock compression ratio of 4, ($V_{\mbox{sh}} \sim 5200 \,$ km s$^-1$; Vink et al. 1998). This value agrees with X-ray Doppler shift measurements, which imply a velocity relative to the shock of about $1400 \,$ km s$^-1$ \citep{wil02}.

We note that we neglected energy loss due to adiabatic expansion, because it is expected to have little effect on the distribution of particles in thin filaments. Similarly, energy losses due to Bremsstrahlung radiation and inverse Compton are not considered, since the synchrotron loss is expected to dominate. To derive analytical solutions, we approximated the synchrotron radiative power for an electron as
\begin{equation} P_{\nu}(\gamma)=(\sigma_{T} c B^2 \gamma^2/6\pi)\delta(\nu-\nu_c), \end{equation}
where $\nu_c=3qB\gamma^2/4\pi mc=3\nu_L \gamma^2 /2$, and $\nu_L$ is the Larmor frequency; here, $\gamma$ is the Lorentz factor of the particle, \emph{m} is its mass, \emph{q} is its charge, and \emph{B} is the magnitude of the magnetic field. We will discuss the effects of this approximation in Section 5.

We assumed that all of the emission originates behind the shock, where the magnetic field is believed to be amplified with respect to the ambient field. The diffusion coefficient was taken as \begin{equation}D(\gamma)=\kappa\frac{m c^{3}\gamma}{3qB},\end{equation} where $\kappa$ is a proportionality constant to be determined. The case when $\kappa=1$ is referred to as Bohm diffusion. Other types of diffusion are also being studied, including Kolmogorov and Kraichnan turbulences \citep{kol41,kra65}, but will be discussed in detail elsewhere (D. Lomiashvili et al. 2010, in preparation).

There are four main parameters in the model: magnetic field, spectral index of electrons ($\Gamma$), diffusion length ($l_{\mbox{dif}}$) and advection length ($l_{\mbox{adv}}$). The diffusion and advection lengths are defined as \begin{equation} l_{\mbox{dif}} = \left(\frac{\kappa m c^3}{\psi q B^3}\right)^{1/2},\end{equation}  \begin{equation}l_{\mbox{adv}}(\gamma) = \frac{V_{\mbox{adv}}}{\psi B^2 \gamma},\end{equation}
respectively, where $\psi=\sigma_{T}/6 \pi mc$. For convenience, we combined these quantities to define two new parameters, $\Lambda_{\mbox{dif}} = l_{\mbox{dif}}/R$ and $\zeta = l_{\mbox{adv}}(1 keV)/l_{\mbox{adv}}$, which can be determined from the data.

\section{Results}
We implemented the model in XSPEC as a table model and applied it to the spectra of the filaments. For each spectral fit, the line features and the hydrogen column density were fixed to values found in the corresponding power-law fits. The quality of the model fits is the same as that of the fits by this phenomenological (power-law) model.

\subsection{Magnetic field and diffusion coefficient}
From the best-fit $\Lambda_{\mbox{dif}}$ and $\zeta$, we derived the magnetic field and diffusion coefficient for each filament (Equations (3) and (4)):
\begin{equation}B \approx 56\mu \mbox{G}\, \left(\frac{\Lambda_{\mbox{dif}}}{0.02}\right)^{-2/3} \left(\frac{\zeta}{5.0}\right)^{-2/3} \left(\frac{V_{\mbox{adv}}}{1.3 \times 10^{8}\,cm\,s^{-1}}\right)^{2/3} \left(\frac{R}{10^{19} \,cm}\right)^{-2/3},\end{equation} \begin{equation}\kappa \approx 0.05 \left(\frac{\zeta}{5.0}\right)^{-2} \left(\frac{V_{\mbox{adv}}}{1.3 \times 10^{8} \,cm \,s^{-1}}\right)^2.\end{equation} \\
The results are summarized in Table \ref{tbl-3}. The indices of the injected electron spectrum found vary from 2.6 to 4 and the magnetic field ranges from $\sim$ 30 $\mu$G to 70 $\mu$G, while $\kappa$ stays around $0.02$ (but can reach up to $0.1 - 0.15$). We should stress, however, that the error bars for the last two quantities are considerable. The magnetic fields for Filaments 1 and 7 are most uncertain, due to the difficulty in constraining $\zeta$.

\subsection{Maximum energy of electrons}
The accelerated electrons will lose their energy due to synchrotron radiation. The evolution of the particle's Lorentz factor, $\gamma$, is given by \begin{equation}\left(\frac{1}{\gamma}\frac{d\gamma}{dt}\right)_{\mbox{loss}}= -\psi B^{2} \gamma \, .\end{equation}

A maximum energy will be reached by the particle when this loss becomes equal to the acceleration rate. If we assume that the mean magnetic field is perpendicular to the shock normal, then for a compression ratio of 4 we can write the acceleration rate as \citep{jok87} \begin{equation}\left(\frac{1}{\gamma}\frac{d\gamma}{dt}\right)_{\mbox{acc}}=\frac{V_{\mbox{sh}}^{2}}{32 \kappa D_{B}} \, ,\end{equation}
where $D_B=r_gc/3$ is the Bohm diffusion coefficient and $r_g=(mc^{2}/qB)\gamma$ is the particle gyroradius.

Our assumption about the direction of the magnetic field is justified by our estimates of the diffusion coefficient, which constrain the obliquity angle to be nearly $90^{\circ}$ (see below).

The maximum energy for an electron then is given by \begin{equation}E_{\mbox{max}} \approx (660\, TeV) \left(\frac{\kappa}{0.05}\right)^{-1/2} \left(\frac{B}{30\,\mu \mbox{G}}\right)^{-1/2} \left(\frac{V_{\mbox{sh}}}{5.2\times10^{8}\,cm\,s^{-1}}\right).\end{equation}

\subsection{Shock obliquity and turbulence level}
We considered diffusion in the radial direction, with a corresponding diffusion coefficient given by \citep{jok87,bla87}: \begin{equation} D=D_{\parallel}\cos^2\theta+D_{\perp}\sin^2\theta \, , \end{equation}
where $\theta$ is the angle between the mean magnetic field and the normal direction of the shock.
Here, we assumed that the kinetic theory relations, $D_{\parallel}=\eta D_{B}$ for the diffusion coefficient along the mean direction of the field and $D_{\perp}=\eta D_{B}/(1+\eta^2)$ for the component of the diffusion coefficient perpendicular to the mean direction of the field, hold \citep[e.g.,][]{for74}. In these equations, $\eta \equiv \lambda_{\mbox{mfp}}/r_g$ is the particle's gyrofactor, which is the ratio of the scattering mean free path, $\lambda_{\mbox{mfp}}$, to the particle gyroradius, $r_g$ \citep{hay69,mel80}. Since isotropic Bohm-type diffusion is assumed here, we can rewrite Equation (2) in the form $D=\kappa D_B$. From Equation (10) we have \begin{equation}\kappa=\eta \left(\cos^2\theta+\frac{\sin^2\theta}{1+\eta^2}\right).\end{equation}

When diffusion is taken as a perturbation in the particle orbits, the fraction $\eta$ can be written in terms of the energy content in the resonant MHD waves \citep[e.g.,][]{bla87}, $\eta=(\delta B/B)^{-2}$, of amplitude $\delta B$. We can then use Equation (11) to constrain $\theta$ and the level of turbulence. For most cases, we found that $\kappa \approx 0.02$ which requires that $86^{\circ} \le \theta \le 90^{\circ}$ and $6 \le \eta \le 16$. This implies a relatively high turbulence level, $$0.25\le \frac{\delta B}{B} \le 0.4 \, .$$

\subsection{Forward shock and Precursor}
\begin{figure}[htp]
\centering
\includegraphics[width=12cm]{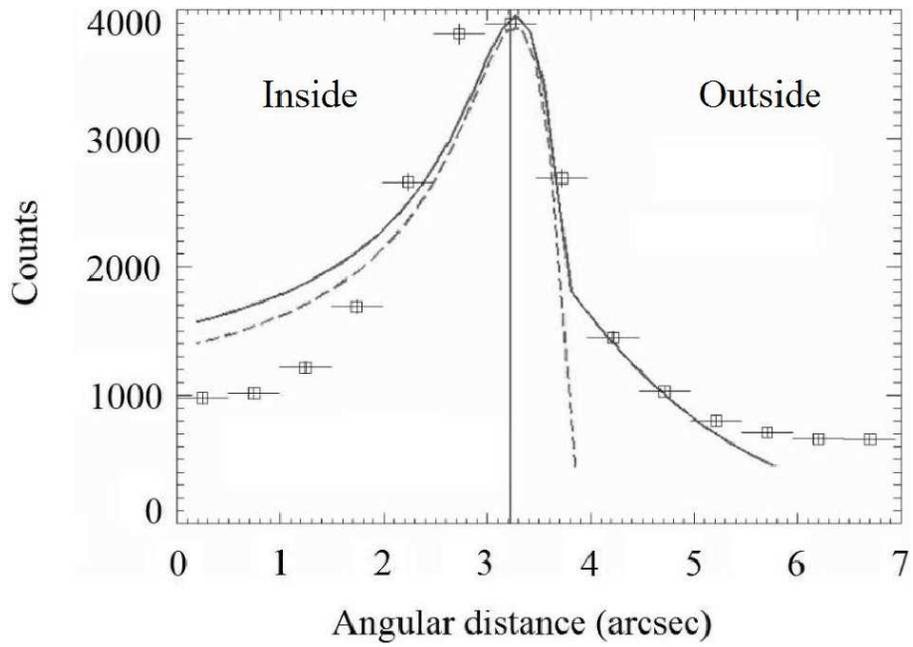}
\caption{Fit to the linear profile of Filament 5 (also see Figure 2) obtained by the model described in Section 4 (dashed line) and calculated with the addition of a precursor (solid line). \label{fig-5}}
\end{figure}

The results indicate that most of the radiation is originated from behind the forward shock (see Figure \ref{fig-2}). However, the model could not explain the observed linear profile of the filaments (see Figure \ref{fig-5} for an example). The model predicts a sharp decline after the peak, which is not observed.

We speculated that some of the emission may come from a precursor \citep{ell94}. We estimated the contribution from the precursor by requiring that the distribution function should be continuous across the shock. The precursor would consist of particles that have diffused across the shock but remain energetic. Specifically, it evolves in the presence of a magnetic field consistent with that of the un-shocked medium surrounding the SNR, here assumed to be 4 times lower than the compressed field estimated downstream; however, we assumed that $\kappa$ remains the same.

Addition of this component substantially improves the predicted profile shape, as shown in Figure \ref{fig-5}. On the other hand, we found that the inclusion of a precursor hardly affects the spectral parameters. More details will be discussed in a future publication (D. Lomiashvili et al. 2010, in preparation).

\section{Discussion}
From the power-law fits to the spectra of the filaments in Cas A, it is seen that the emission from the inner regions is consistently softer, by about 10\%, than that from the outer regions. This seems to be consistent with the effect expected from radiative cooling, since the outer regions have had less time since they interacted with the shock. When only synchrotron losses and advection are taken into account, however, the predicted difference between the inner and outer photon indices is the same in all filaments. The data show that this difference can change from one filament to another.

Also, from a consideration of synchrotron losses, one might expect that the widths of the filaments get narrower at higher energies. In fact, if synchrotron cooling and advection were the only processes controlling the plasma distribution, the width of a non-thermal filament can be estimated as $\emph{w} = {V}_{\mbox{adv}}{\tau}_{s}$, or the distance the particles are advected before radiating away their energy. This can also be written as $w = V_{\mbox{adv}}/{\psi}{B}^{2}{\gamma}$, where $\psi=\sigma_T/6 \pi mc$, with $\sigma_T$ the Thompson cross section for electrons, $B$ the magnetic field, and $\gamma$ the Lorentz factor of the accelerated particle, which when assuming emission peaked at the Larmor frequency $\nu_{L}$ can be written as $(\nu/\nu_{L})^{1/2}$. Therefore, an important dependence of the widths on the frequency of the radiation, of the form $\emph{w} \,\, \alpha \,\, \nu^{-1/2}$, would result. However, the data suggest that no dependence exists.

These observations seem to point at the existence of additional mechanisms affecting the evolution of the plasma distribution and are found to be consistent with the model used. In this model, the difference between the photon indices of the inner and outer regions is regulated by diffusion, and it is determined mostly by the ratio of advection to diffusion lengths, $\zeta$, whereas it is found that varying $\Lambda_{\mbox{dif}}$ ($\equiv l_{\mbox{dif}}/R$) produces changes mainly in the calculated width of the non-thermal filaments without considerably affecting the model spectra. This was also seen when carrying out the fits, since the values of $\chi^2$ did not change appreciably for a wide range of values of $\Lambda_{\mbox{dif}}$, and therefore additional constraints on this parameter were necessary. As a constrain, we used the values of $\Lambda_{\mbox{dif}}$ that were calculated to match the filamentary widths at half intensity to the actual data as initial values for the fits. This parameter ranges from 0.014 to 0.034, while $\zeta$ varies from 3.3 to 8.9.

Due to the role that $\zeta$ plays in the model, it should be possible to correlate the spatial differences between the photon indices with the value of the proportionality constant in the diffusion coefficient, $\kappa$. It is seen that Filaments 1, 6, and 7 show the highest values for $\kappa$ (although considerable uncertainties were obtained) and that the relative change in photon spectral index from inner to outer regions is the lowest for these filaments (although Filament 5 shows a similar change). The amount of diffusion can change 1 order of magnitude for the different filaments depending on the degree of spatial spectral variation observed. It can be argued that the diffusion of particles tends to homogenize the plasma distribution and lower the difference between the inner and outer photon indices. This was also seen in the simulations where higher values of $\kappa$ were used.

The particle spectral indices are found close to 3, although steeper values are also seen (up to 4 for Filament 6). This index corresponds to the power-law distribution of electrons resulting from shock acceleration processes. After the particles evolve in the magnetic field, one might expect to see steeper spectral indices, especially at higher X-ray energies. Such steep distributions might have been seen before. For instance, when comparing the 10-32 keV \emph{RXTE} Proportional Counter Array spectrum of Cas A with the predicted synchrotron emission spectrum dominating the band from 0.3 keV to 7 keV, the observed excess can be accounted for by an additional contribution of non-thermal Bremsstrahlung radiation from a steep power-law (index $\sim$4.1) population of electrons interacting with other electrons and with ions \citep{all08}.

Besides accounting for these observations, we estimated the magnetic field in each filament as well as the level of turbulence. The derived magnetic field is on average below the values previously inferred for the NE rim (Filaments 1-3) by assuming that the width of the filament (taken as $1^{\prime\prime}.5 - 4^{\prime\prime}$) is determined by synchrotron losses and advection only \citep{vink03}, $B_{\mbox{sync}}\sim 80 - 160 \mu$G. The difference might be explained by the fact that our estimates take into account these two processes but additionally consider diffusion. The turbulent magnetic field is constrained to be $0.25\le \frac{\delta B}{B} \le 0.4$. Moreover, the ordered field is found to be nearly perpendicular to the shock front ($86^{\circ} \le \theta \le 90^{\circ}$), consistent with an expansion inside a toroidal external field produced originally by the progenitor star. The inferred magnetic field in the filaments is still higher than that expected from magnetic field amplification of the interstellar field ($\sim 3 \mu$G), as was also pointed out by Vink $\&$ Laming. Perhaps the interstellar field surrounding Cas A is higher, or the field has been amplified by the high-energy particles near the shock front through nonlinear wave growth.

In some cases, the magnetic field might be much higher in other filaments of Cas A \citep{ato00}. \citet{pat07} and \citet{uch08} have observed X-ray variability of some of the non-thermal filaments seen in projection near the reverse shock on a time scale of a few years. Assuming that this variability is related to synchrotron cooling and DSA, and that the diffusion is close to the Bohm limit with the field parallel to the shock front, Uchiyama $\&$ Aharonian estimated that the field required would be $\sim 1\,$mG. However, there is still debate as to whether the observed variability is associated with filaments in the forward shock, and there are other plausible scenarios for explaining it \citep[e.g.,][]{byk08}.

Models describing the non-equilibrium ionization behind an SNR forward shock recently developed \citep{ell07,pat09b} show that efficient DSA could increase the ionization fraction of some elements. In their model, Patnaude et al. (2009) have found that ionization occurs more rapidly and closer to the shock when the particles experience efficient shock acceleration, meaning that spatial variations in the intensity of the thermal emission near the shock might be used to probe the properties of cosmic rays. We looked at the spectral features and estimated the equivalent widths of Si K XIII and S K XV in order to quantify any possible spatial variations in their intensity. In Table \ref{tbl-4}, we summarize the results. Unfortunately, the error bars associated with the equivalent widths that we observed are too large for us to draw definitive conclusions.

Finally, our model assumes that synchrotron radiation is monochromatic at $\nu_{c}$ (see Equation (1)), which is clearly an oversimplification. When we incorporated the full synchrotron spectrum into the model, we saw changes in the model parameters. For instance, a fit to the spectrum of Filament 5 with the revised model leads to $B\approx70\mu$G and $\kappa\approx0.06$, which are different to the values shown in Table \ref{tbl-3}. However, the changes do not qualitatively modify our conclusions. The obliquity angle is still close to $90^{\circ}$ and the turbulent field is $\frac{\delta B}{B}\approx 0.2$. The details of the full model will be presented in a future publication.

\section{Summary}
We summarize our main results as follows.

\begin{enumerate}
\item Spectral evolution is seen across non-thermal filaments in Cas A, with the spectra of the outer regions being harder by about 10\% on average.
\item The widths of the filaments show no significant dependence on photon energy.
\item To account for the observational results (1 and 2), we needed to include the effects of diffusion. If we restrict to Bohm-type diffusion, we could quantify the level of turbulence ($0.25\le \frac{\delta B}{B} \le 0.4$) as well as the diffusion itself ($\kappa \approx 0.02 - 0.15$). Moreover, we found that the magnetic field is of the order of tens of $\mu$G, varying from filament to filament, and that the field is nearly perpendicular to the shock front.
\end{enumerate}

These results are in overall agreement with models of cosmic ray acceleration in the shocks of SNRs. They imply that there is a high level of magnetic turbulence in the non-thermal filaments associated with the forward shock of Cas A, as well as magnetic field amplification. Both of these conditions are necessary to efficiently accelerate cosmic rays.

Regarding the shock orientation, our analysis of a sampling of non-thermal filaments, which have good azimuthal coverage of the remnant, implies that the obliquities are close to $90^{\circ}$, which is consistent with the expansion of Cas A in the wind environment produced by the progenitor \citep{che03}.

Efficient cosmic ray acceleration in the shock of Cas A would have implications regarding the acceleration of protons (and heavier ions), which may interact with cold ambient protons and produce neutral pions that would decay into gamma rays, leaving a signature in the spectrum of the remnant in the GeV to TeV energy range. This signature could in principle be detected.

\acknowledgments
We thank M. Laming, M. Pohl, and S. Reynolds for useful discussions. This research has made use of data obtained from the \emph{Chandra Data Archive} and the \emph{Chandra Source Catalog}, and software provided by the \emph{Chandra X-ray Center (CXC)} in the application package CIAO. This work has also made use of NASA's Astrophysical Data System. We gratefully acknowledge financial support from NASA and Purdue University.

\pagebreak

\begin{table}
\begin{center}
\caption{Widths of Linear Profiles of Non-Thermal Filaments. \label{tbl-1}}
\begin{tabular}{|c|c|c|c|}
\tableline\tableline
Filament &\multicolumn{1}{c|}{0.3 - 2.0 keV}& \multicolumn{1}{c|}{3.0 - 6.0 keV}&\multicolumn{1}{c|}{6.0 - 10.0 keV}\\
\tableline
1 &1.58$^{+0.07}_{-0.06}$&1.23$^{+0.07}_{-0.07}$&1.24$^{+0.41}_{-0.24}$ \\
2 &1.49$^{+0.13}_{-0.11}$&1.07$^{+0.11}_{-0.10}$&1.25$^{+0.37}_{-0.27}$ \\
3 &1.46$^{+0.09}_{-0.07}$&1.03$^{+0.07}_{-0.07}$&1.08$^{+0.41}_{-0.28}$ \\
4 &2.34$^{+0.26}_{-0.20}$&1.64$^{+0.26}_{-0.18}$& 1.65$^{+3.9}_{-0.63}$ \\
5 &0.95$^{+0.03}_{-0.02}$&0.78$^{+0.03}_{-0.02}$&0.77$^{+0.14}_{-0.12}$ \\
6 &1.28$^{+0.07}_{-0.06}$&0.98$^{+0.08}_{-0.07}$&1.25$^{+6.1}_{-0.43}$ \\
7 &1.35$^{+0.08}_{-0.06}$&1.08$^{+0.08}_{-0.07}$&1.26$^{+0.98}_{-0.34}$ \\
8 &1.10$^{+0.17}_{-0.06}$&0.92$^{+0.06}_{-0.05}$&0.92$^{+0.35}_{-0.20}$ \\
9 &2.01$^{+0.24}_{-0.18}$&1.50$^{+0.16}_{-0.14}$&1.74$^{+3.3}_{-0.51}$ \\
\tableline
\end{tabular}
\tablecomments{Error intervals at the 90\%\ confidence level and widths are in arcseconds.}
\end{center}
\end{table}

\begin{table}
\begin{center}
\caption{Best-Fit Spectral Power Laws. \label{tbl-2}}
\begin{tabular}{|c|c|c|c|c|c|c|}
\tableline\tableline
&\multicolumn{3}{c|}{Inside}&\multicolumn{3}{c|}{Outside}\\ \cline{2-7}
\raisebox{0.1in}[0in][0in]{Filament} &\multicolumn{1}{c}{$\gamma$\tablenotemark{a}} & \multicolumn{1}{c}{$\mbox{Norm}$\tablenotemark{b}} & \multicolumn{1}{c|}{$\chi^2/\mbox{dof}\tablenotemark{c}$} & \multicolumn{1}{c}{$\gamma\tablenotemark{a}$}&\multicolumn{1}{c}{$\mbox{Norm}$\tablenotemark{b}} & \multicolumn{1}{c|}{$\chi^2/\mbox{dof}\tablenotemark{c}$}\\
\tableline
1 &2.49$^{+0.12}_{-0.09}$&4.58 &0.45 &2.41$^{+0.15}_{-0.13}$& 2.83 &0.45 \\
2 &2.59$^{+0.2}_{-0.16}$&2.01 &0.61 &2.16$^{+0.11}_{-0.1}$& 3.61 &0.34 \\
3 &2.51$^{+0.09}_{-0.08}$&6.85 &0.46 &2.16$^{+0.11}_{-0.10}$& 3.38 &0.42 \\
4 &3.0$^{+0.22}_{-0.21}$&2.54 &0.84 &2.50$^{+0.13}_{-0.11}$& 3.58 &0.45 \\
5 &2.71$^{+0.15}_{-0.13}$&3.98 &0.40 &2.55$^{+0.15}_{-0.08}$& 5.21 &0.45 \\
6 &3.18$^{+0.18}_{-0.11}$&8.19 &0.36 &2.99$^{+0.13}_{-0.16}$& 6.07 &0.43 \\
7 &2.64$^{+0.12}_{-0.10}$&5.22 &0.35 &2.49$^{+0.18}_{-0.16}$& 2.46 &0.50 \\
8 &2.89$^{+0.15}_{-0.14}$&7.33 &0.42 &2.61$^{+0.18}_{-0.17}$& 5.48 &0.43 \\
9 &2.51$^{+0.26}_{-0.16}$&4.77 &0.43 &2.31$^{+0.43}_{-0.26}$& 1.98 &0.51 \\
\tableline
\end{tabular}
\tablecomments{Error intervals at the 90\%\ confidence level}
\tablenotetext{a}{Power-law index}
\tablenotetext{b}{Power-law normalization, in units of $10^{-5}$ photons keV$^{-1}$ cm$^{-2}$ s$^{-1}$}
\tablenotetext{c}{Reduced chi-squared of the fit}
\end{center}
\end{table}

\begin{table}
\begin{center}
\caption{Key Parameters of the Diffusion-Advection Model. \label{tbl-3}}
\begin{tabular}{|c|c|c|c|}
\tableline\tableline
Filament & \multicolumn{1}{c|}{$B (\mu\mbox{G}) $} & \multicolumn{1}{c|}{$\kappa $} & \multicolumn{1}{c|}{$\Gamma $} \\
\tableline
1 &72$^{+24}_{-70}$ &0.12$^{+0.2}_{-0.1}$ &2.7$^{+0.2}_{-0.06}$ \\
2 &37$^{+10}_{-11}$ &0.02$^{+0.01}_{-0.018}$ &2.63$^{+0.3}_{-0.1}$ \\
3 &53$^{+10}_{-18}$ &0.02$^{+0.01}_{-0.015}$ &2.66$^{+0.04}_{-0.15}$ \\
4 &40$^{+15}_{-5}$ &0.02$^{+0.02}_{-0.003}$ &3.6$^{+0.2}_{-0.3}$ \\
5 &52$^{+26}_{-31}$ &0.025$^{+0.04}_{-0.02}$ &3.2$^{+0.1}_{-0.1}$ \\
6 &56$^{+20}_{-30}$ &0.1$^{+0.08}_{-0.08}$ &4.0$^{+0.1}_{-0.2}$ \\
7 &66$^{+40}_{-60}$ &0.15$^{+0.2}_{-0.1}$ &3.0$^{+0.1}_{-0.15}$ \\
8 &35$^{+16}_{-19}$ &0.02$^{+0.02}_{-0.01}$ &3.5$^{+1.4}_{-0.4}$ \\
9 &29$^{+10}_{-14}$ &0.02$^{+0.02}_{-0.015}$ &2.6$^{+0.06}_{-0.1}$ \\
\tableline
\end{tabular}
\tablecomments{Error intervals at the 90\%\ confidence level}
\end{center}
\end{table}

\begin{table}
\begin{center}
\caption{Equivalent Widths of Si K XIII and S K XV Emission Lines \label{tbl-4}}
\begin{tabular}{|c|c|c|c|c|}
\tableline\tableline
&\multicolumn{2}{c|}{Inside}&\multicolumn{2}{c|}{Outside}\\ \cline{2-5}
\raisebox{0.1in}[0in][0in]{Filament} &\multicolumn{1}{c}{$\emph{Si K XIII}$} & \multicolumn{1}{c|}{$\emph{S K XV}$} & \multicolumn{1}{c}{$\emph{Si K XIII}$}&\multicolumn{1}{c|}{$\emph{S K XV}$}\\
\tableline
1 &130$^{+90}_{-57}$&142$^{+120}_{-100}$ &169$^{+140}_{-95}$ &191$^{+190}_{-160}$\\
2 &175$^{+168}_{-102}$&188$^{+225}_{-185}$ &99.1$^{+98}_{-75}$ &137$^{+157}_{-130}$\\
3 &96.2$^{+64}_{-50}$&134$^{+109}_{-102}$ &97.5$^{+100}_{-59}$ &149$^{+170}_{-149}$\\
4 &177$^{+144}_{-109}$&170$^{+228}_{-168}$ &99.6$^{+100}_{-75}$ &145$^{+160}_{-140}$\\
5 &97.4$^{+94}_{-67}$&173$^{200}_{-155}$ &60.0$^{+70}_{-59}$ &133$^{+130}_{-120}$\\
6 &188$^{+115}_{-70}$&203$^{+142}_{-99}$ &173$^{+150}_{-81}$ &190$^{+200}_{-120}$\\
7 &85.1$^{+92}_{-60}$&141$^{+150}_{-140}$ &104$^{+140}_{-97}$ &118$^{+220}_{-116}$\\
8 &$98.5^{+150}_{-68}$&94.0$^{+200}_{-92}$ &68.5$^{+290}_{-68}$ &75.5$^{+370}_{-75.5}$\\
9 &133$^{+148}_{-79}$&95.2$^{+205}_{-95}$ &153$^{+400}_{-108}$ &91.1$^{420}_{-90}$\\
\tableline
\end{tabular}
\tablecomments{Error intervals at the 90\%\ confidence level}
\end{center}
\end{table}
\clearpage
\pagebreak

\end{document}